\documentstyle[11pt,aaspp4,psfig]{article}  

\def\ms{$M_\odot$}
\def\e#1{$\times$ 10$^{#1}$}
\def\etal{et al. }
\def\ni{$^{56}$Ni}
\def\co{$^{56}$Co}

\def\kms{km s$^{-1}$}
\def\pa{\partial}
\def\lsim{\mathrel{\rlap{\lower 4pt \hbox{\hskip 1pt $\sim$}}\raise 1pt \hbox
        {$<$}}}
\def\gsim{\mathrel{\rlap{\lower 4pt \hbox{\hskip 1pt $\sim$}}\raise 1pt \hbox
        {$>$}}}

\begin{document}

\title{The Peculiar Type Ic Supernova 1997ef: Another Hypernova} 

\author{Koichi IWAMOTO$^1$, Takayoshi NAKAMURA$^2$, 
Ken'ichi NOMOTO$^{2,3}$, Paolo A. MAZZALI$^{3,4}$, \\I. John DANZIGER$^4$,
Peter GARNAVICH$^{5}$, Robert KIRSHNER$^{5}$, Saurabh JHA$^{5}$, \\
David BALAM$^{6}$, and John THORSTENSEN$^{7}$}

\affil{$^{1}$ Department of Physics,
College of Science and Technology, Nihon University,
Chiyoda-ku, Tokyo 101-8308, Japan \\
$^{2}$ Department of Astronomy, School of Science, University
of Tokyo, Bunkyo-ku, Tokyo 113-0033, Japan \\
$^{3}$ Research Center for the Early Universe, School of Science,
University of Tokyo, Bunkyo-ku, Tokyo 113-0033, Japan \\
$^{4}$ Osservatorio Astronomico di Trieste, via G.B.
Tiepolo 11, I-34131 Trieste, Italy \\
$^{5}$ Harvard-Smithsonian Center for Astrophysics, 
60 Garden Street, Cambridge, MA 02138, USA \\
$^{6}$ University of Victoria, Victoria, BC, Canada \\
$^{7}$ Dartmouth College, Department of Physics and Astronomy,
Hanover, NH 03755, USA \\}

\received{May 26, 1998 (in original form); October 20, 1999}
\accepted{December 17, 1999}

\vskip 0.5cm

\centerline{"To appear in the Astrophysical Journal, Vol. 534 (2000)"}

\begin{abstract}

SN 1997ef has been recognized as a peculiar supernova from its light
curve and spectral properties. The object was classified as a Type Ic
supernova (SN Ic) because its spectra were dominated by broad
absorption lines of oxygen and iron, lacking any clear signs of
hydrogen or helium line features.  The light curve is very different
from that of previously known SNe Ic, showing a very broad peak and a
slow tail.  The strikingly broad line features in the spectra of
SN~1997ef, which were also seen in the hypernova SN 1998bw, suggest
the interesting possibility that SN 1997ef may also be a hypernova.

The light curve and spectra of SN 1997ef were modeled first with a
standard SN~Ic model assuming an ordinary kinetic energy of explosion
$E_{\rm K} = 10^{51}$ erg.  The explosion of a CO star of mass $M_{\rm
CO} \approx 6 M_\odot$ gives a reasonably good fit to the light curve
but clearly fails to reproduce the broad spectral features.  Then,
models with larger masses and energies were explored.  Both the light
curve and the spectra of SN~1997ef are much better reproduced by a C+O
star model with $E_{\rm K} =$ 8 \e{51} erg and $M_{\rm CO} = 10
M_\odot$.  Therefore, we conclude that SN~1997ef is very likely a
hypernova on the basis of its kinetic energy of explosion.  Finally,
implications for the deviation from spherical symmetry are discussed
in an effort to improve the fits to the observations.

\end{abstract}


\section{INTRODUCTION}

\begin{figure}
\centerline{\psfig{figure=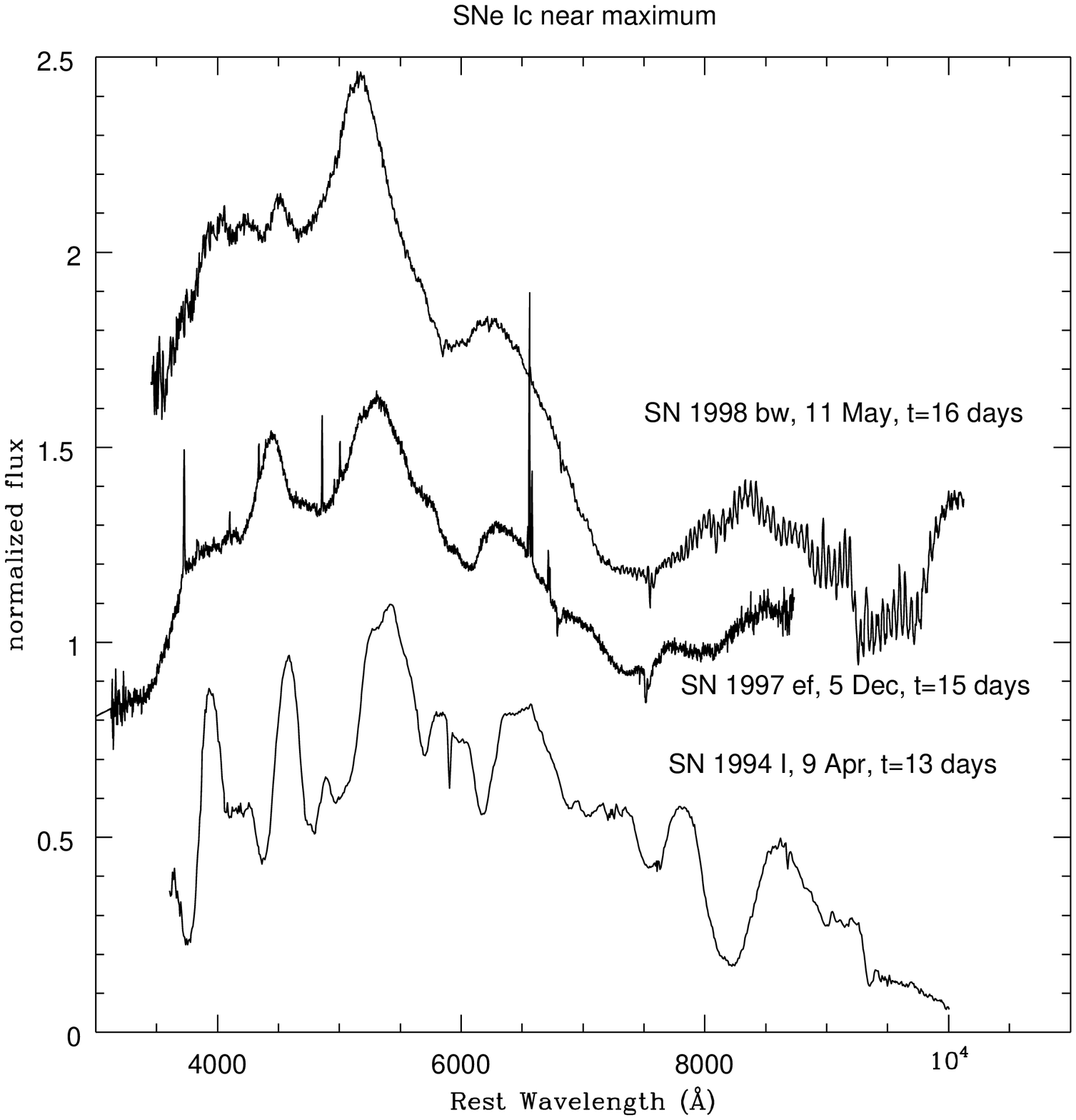,width=12cm}}
\caption{Observed spectra of Type Ic supernovae 1998bw, 1997ef, and 1994I.
\label{spic2}}
\end{figure}

The supernova 1997ef (SN 1997ef) was discovered on November 25, 1997
at an R magnitude of 16.7 near the spiral galaxy UGC4107 (Sano
1997). The first spectrum was taken on November 26 (Garnavich \etal
1997a). Subsequently, photometric and spectroscopic follow-ups have
provided high quality optical light curves and spectra (Garnavich
\etal 1997a,b,c; Hu \etal 1997; Filippenko \etal 1997; Wang \etal
1998).  As seen in Figure \ref{spic2}, the spectra of SN 1997ef are
dominated by broad oxygen and iron lines but do not show any clear
feature of hydrogen or helium (Garnavich \etal 1997b; Filippenko \etal
1997), showing the overall similarity to other Type Ic supernovae (SNe
Ic) 1994I and 1998bw.  This led us to classify SN~1997ef as a SN Ic.

\begin{figure}
\centerline{\psfig{figure=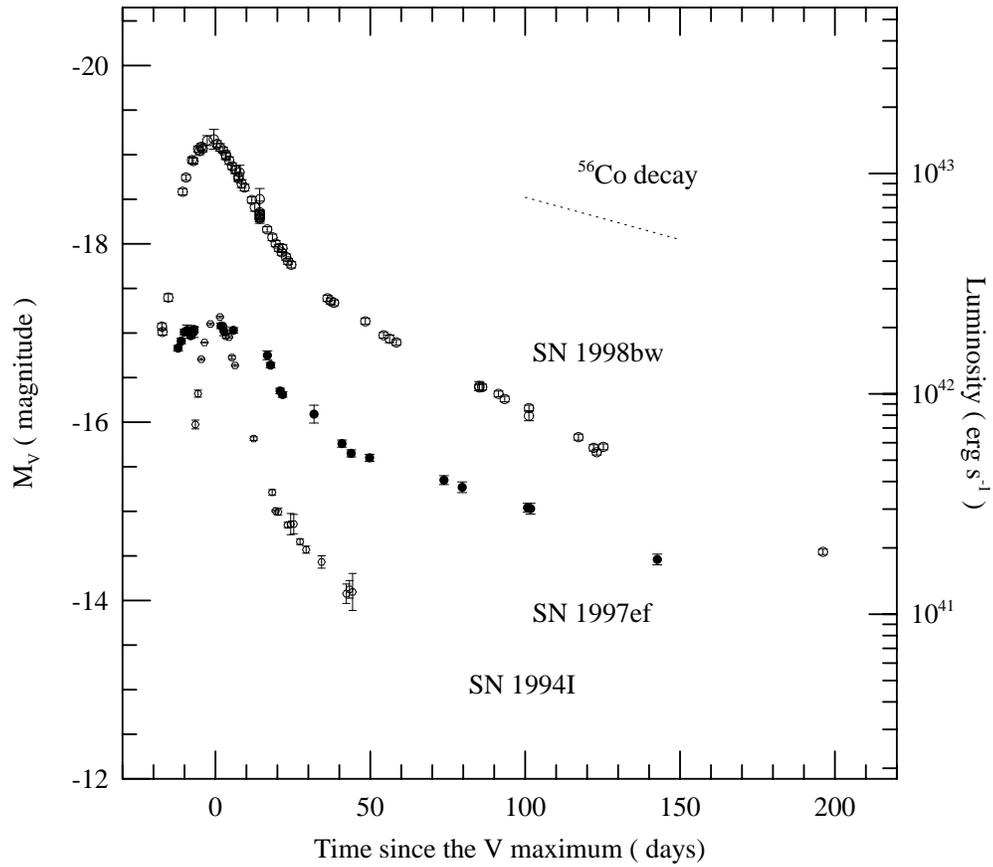,width=15cm}}
\caption{Absolute magnitudes of Type Ic supernovae: the ordinary SN~Ic
1994I (Richmond \etal 1996a, b), the hypernova SN~1998bw (Galama et
al. 1998), and the proposed hypernova SN~1997ef.  The dashed line
indicates the \(^{56}\)Co decay rate.  
\label{lcic}}
\end{figure}

In Figure \ref{lcic} the visual light curve of SN 1997ef (Garnavich
\etal 1997b,c) is compared with those of the SN~Ic 1998bw (Galama et
al. 1998) and the ordinary SN~Ic 1994I (Richmond \etal 1996a,b).
Despite the spectral similarity, the light curve of SN~1997ef is quite
different from those of SN~1998bw and SN~1994I.  It has quite a flat
peak, much broader than those of the other SNe Ic.  Besides, the tail
of the light curve of SN~1997ef starts late and the rate of its
decline is much slower than in other SNe Ic.  It is also true that the
light curves are rather diverse, even in this limited number of
samples, implying a range of energies and/or progenitor masses of SN
Ic explosions.

The most striking and peculiar characteristic of SN~1997ef is the
breadth of its line features.  Such broad spectral features were later
recognized to be a distinguishing property of the spectra in SN 1998bw
(Fig. \ref{spic2}).  SN 1998bw was discovered within the error box of
GRB980425 determined by the BeppoSAX satellite, only 0.9 days after
the date of the gamma-ray burst (GRB), and therefore probably related
to this GRB (Galama et al. 1998).  The very broad spectral features
and the light curve shape have led to the conclusion that SN 1998bw
had an extremely large {\sl kinetic} energy of explosion, $E_{\rm K}
\sim$ 3 \e{52} ergs (Iwamoto et al. 1998; Woosley, Eastman, \& Schmidt
1999).  This was one order of magnitude larger than the energy of
typical supernovae, thus SN 1998bw was termed a ``hypernova'' (Iwamoto
et al. 1998).

The spectral similarities between SN 1997ef and SN 1998bw suggest the
interesting possibility that SN 1997ef may also be a hypernova.  In
fact, a possible connection with a GRB has been suggested for
SN~1997ef: GRB971115 appears to be compatible with the supernova in
the position and the time of occurrence (Wang \& Wheeler 1998).  Since
the statistical significance for this case is much weaker than for the
case of SN~1998bw and GRB980425, it is difficult to confirm the
physical association between SN 1997ef and GRB971115.  However, it is
possible at least to clarify whether SN 1997ef is a hypernova or not
by estimating the kinetic energy of explosion through modeling of
light curves and spectra as in the case of SN 1998bw (Iwamoto et
al. 1998; Mazzali 1999). This is exactly the primary purpose of this
paper.

We constructed supernova progenitor models and performed detailed
hydrodynamics and radiation transfer calculations to obtain light
curves and spectra for the explosion models.  The results were
compared with observations of SN 1997ef, to derive its explosion
energy and the ejecta mass, and thus to determine whether SN 1997ef
was an ordinary SN Ic or a hypernova.  Since the light curves of the
other SNe Ic 1994I and 1998bw were successfully reproduced by the
collapse-induced explosion of C+O stars (Nomoto et al. 1994; Iwamoto
et al. 1994; Iwamoto et al. 1998), we adopted C+O stars as progenitor
models for SN 1997ef as well.

This paper consists of six sections including this introduction.
Section 2 describes the ordinary SN Ic model and the hypernova model.
The method and results of our light curve calculations are presented
in \S3. The synthetic spectra are compared with the observations in
\S4. Section 5 is devoted to discussion on various issues including
the SN-GRB connection and possible progenitor scenarios.  Finally, our
conclusions are summarized in \S6.

\section{EXPLOSION MODELS FOR SUPERNOVAE AND HYPERNOVAE}

We construct hydrodynamical models of an ordinary SN Ic and a
hypernova as follows.

(1) In the ordinary SN Ic model(model CO60), a C+O star with a mass
$M_{\rm CO}= 6.0 M_\odot$ (which is the core of a 25 $M_\odot$
main-sequence star) explodes with kinetic energy of explosion $E_{\rm
K} = 1.0 \times 10^{51}$ ergs and ejecta mass $M_{\rm ej} = M_{\rm
CO}-M_{\rm rem} = 4.6 M_\odot$. Here $M_{\rm rem}$ (= 1.4 \ms) denotes
the mass of the compact star remnant (either a neutron star or a black
hole).

(2) In the hypernova model (CO100), a C+O star of $M_{\rm CO}= 10
M_\odot$ is constructed from the 10 \ms\ He star (which has a 8 \ms\
C+O core) by removing the outermost 2 \ms\ of He layer and extending
the C+O layer up to $10 M_\odot$.  This model corresponds to 30 - 35
$M_\odot$ on the main-sequence.  This progenitor goes off with $E_{\rm
K} = 8.0 \times 10^{51}$ ergs and $M_{\rm ej} = 7.6 M_\odot$, i.e.,
$M_{\rm rem}$ = 2.4 \ms.

The hydrodynamics at early phases was calculated by using a Lagrangian
PPM code (Colella \& Woodward 1984) with a simple nuclear reaction
network including 13 alpha elements (M\"uller 1986).  Detailed
post-processing calculations were carried out with a larger-size
nuclear reaction network including 240 isotopes (Hix \& Thielemann
1996).  The explosion is triggered by depositing thermal energy in a
couple of zones just below the mass cut so that the final kinetic
energy becomes the required value. The position of the mass cut is
adjusted for the ejected mass of $^{56}$Ni to be $M(^{56}$Ni) $= 0.15
M_\odot$.

The compact remnant in CO60 is likely a neutron star because $M_{\rm
rem}$ = 1.4 \ms, while it may be a black hole in CO100 because $M_{\rm
rem}$ (= 2.4 \ms) may well exceed the maximum mass of a stable neutron
star.  The above values of $M_{\rm rem}$ are determined so that
$M(^{56}$Ni)$= 0.15 M_\odot$ is ejected to reproduce the maximum
brightness of SN~1997ef by the radioactive decay heating of $^{56}$Ni
and $^{56}$Co.

These model parameters are summarized in Table 1. They can be
constrained by comparing the calculated light curves and synthetic
spectra with observations.  The parameters of models CO21 for SN~1994I
(Nomoto et al.  1994; Iwamoto et al. 1994) and CO138 for SN~1998bw
(Iwamoto et al. 1998) are also given in Table 1.  We constructed the
progenitor model by attaching a thin hydrostatic and
in-thermal-equilibrium C+O envelope to the C+O core of the
presupernova model (Nomoto \& Hashimoto 1988; Hashimoto 1995).

\begin{figure}
\centerline{\psfig{figure=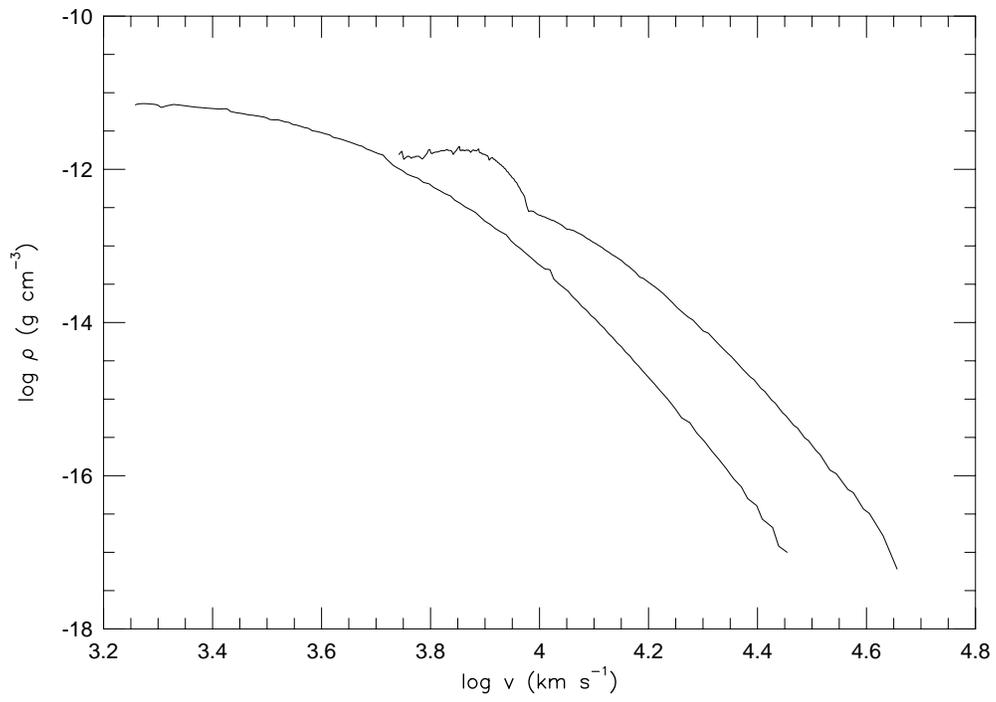,width=15cm}}
\caption{Density distributions against the velocity of
homologously expanding ejecta for CO60 and CO100.
\label{vrhonew}}
\end{figure}

\begin{figure}\
\centerline{\psfig{figure=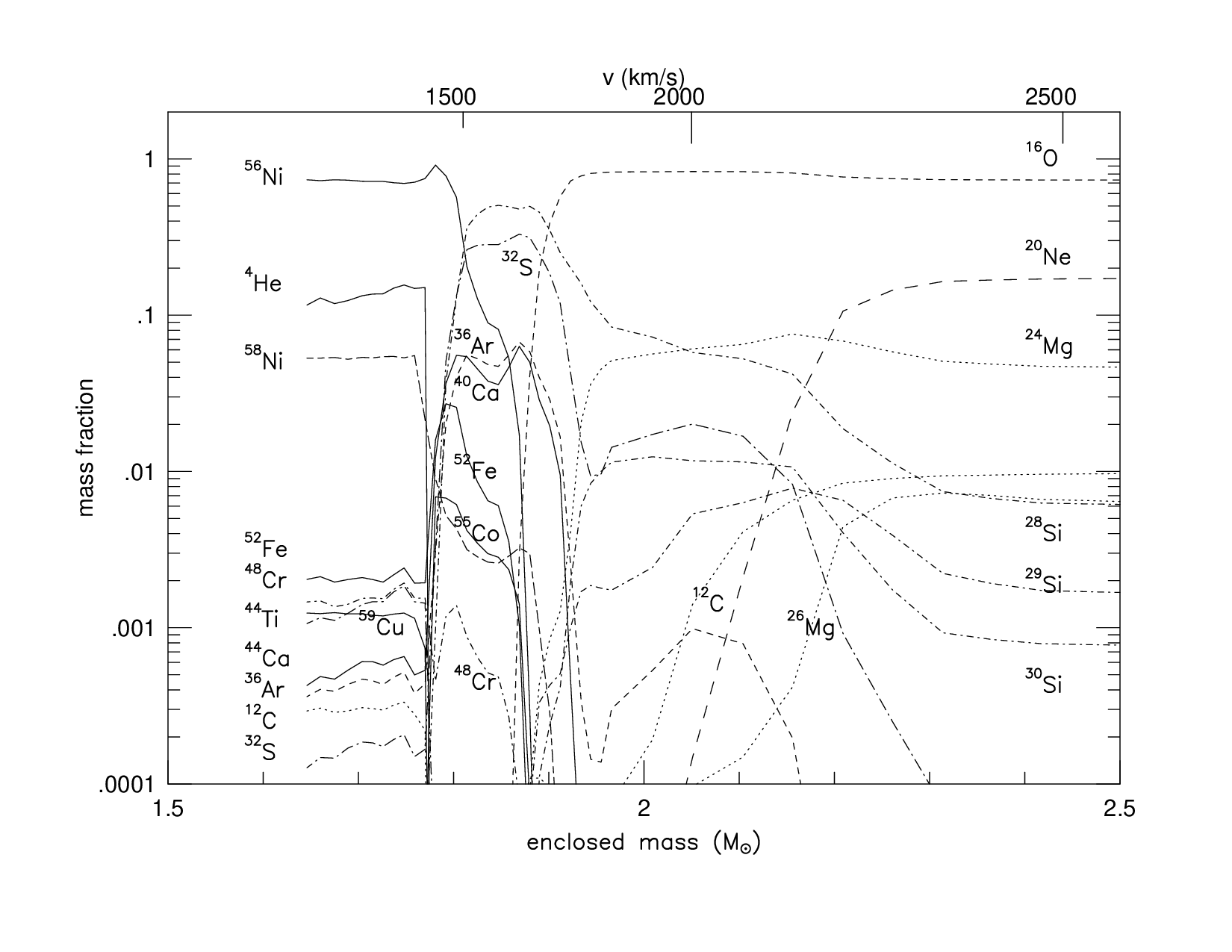,width=15cm}}
\caption{Chemical composition of model CO60 plotted against
the expansion velocity.
\label{nusco60}}
\end{figure}

\begin{figure}
\centerline{\psfig{figure=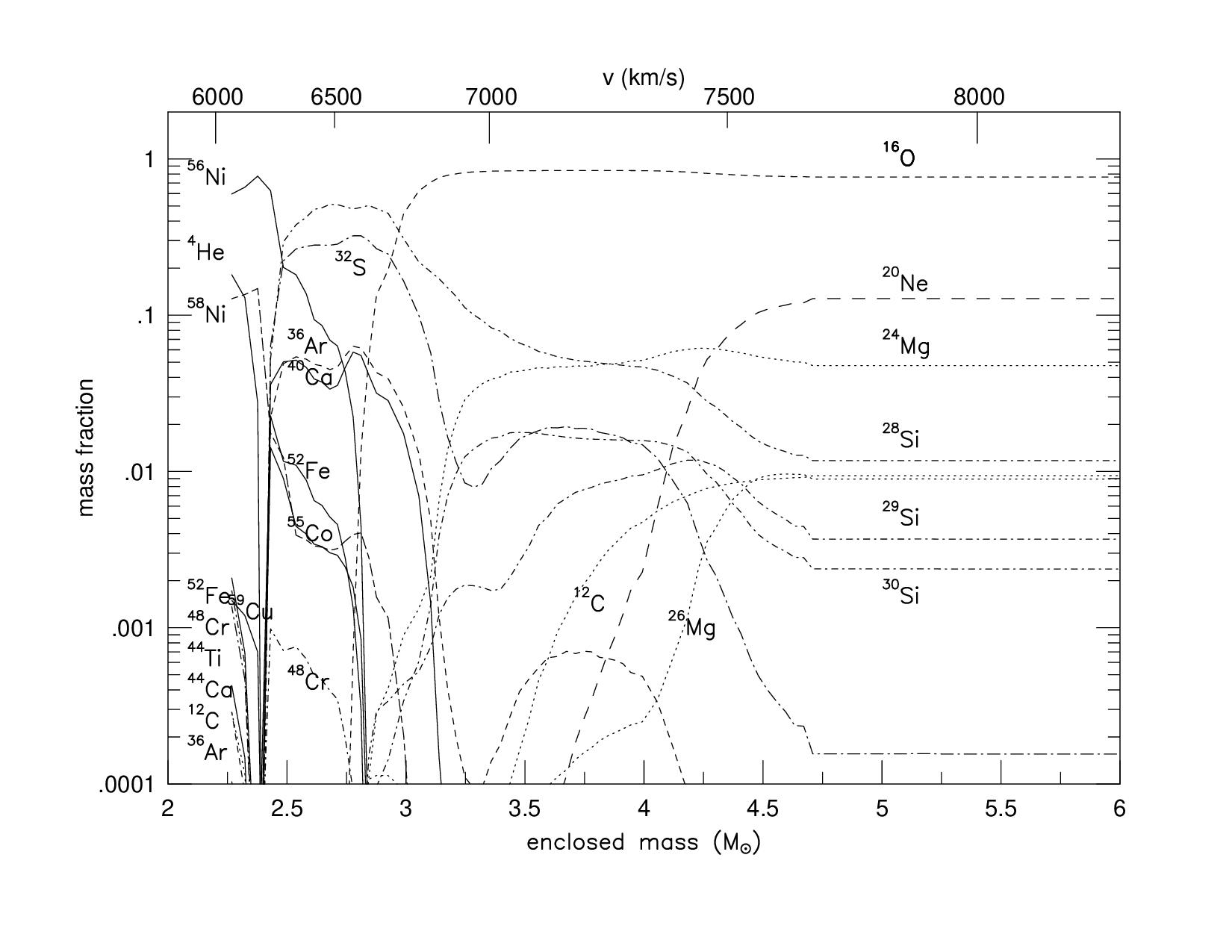,width=15cm}}
\caption{Chemical composition of model CO100 plotted against
the expansion velocity.
\label{nusco100}}
\end{figure}

The expansion soon becomes homologous so that $v \propto r$.  The
solid lines in Figure \ref{vrhonew} show the density distributions in
the velocity space for CO60 and CO100 at $t =$ 16 days.  The expansion
velocities are clearly higher in CO100 than in CO60.  Figures
\ref{nusco60} and \ref{nusco100} show the composition structure of 
models CO60 and CO100, respectively, against the expansion velocity
and the Lagrangian mass coordinate of the progenitor.  In CO100, the
Fe and Si-rich layers expand much faster than in CO60.  The total
amount of nucleosynthesis products are summarized in Table 2.

\section{LIGHT CURVE MODELS}

\subsection{Radiation Hydrodynamics Code}

The light curve calculation is started, when the ejecta have reached
the homologous expansion phase, with a one-dimensional spherically
symmetric radiation transfer code (Iwamoto 1997).

The code solves the multi-frequency radiative transfer equation for
the specific intensity $I_{\nu}$ in the comoving frame, including all
terms up to the first order in $v/c$ (Mihalas \& Mihalas 1984),

\begin{eqnarray}
\frac{1}{c} \frac{\pa I_{\nu}}{\pa t} + 
  \frac{\mu}{r^{2}}\frac{\pa}{\pa r}(r^{2}I_{\nu}) 
&+& \frac{\pa}{\pa \mu} \left\{ (1-\mu^{2}) \left[
 \frac{1}{r} + \frac{\mu}{c} \left( \frac{v}{r}-\frac{\pa v}{\pa r} \right) 
 \right] I_{\nu} \right\}   					   \nonumber\\ 
&-& \frac{\pa}{\pa \nu} \left\{ \nu \left[ (1-\mu^{2}) \frac{v}{cr} 
 + \frac{\mu^{2}}{c}\frac{\pa v}{\pa r} \right] I_{\nu} \right\}   \nonumber\\ 
&+& \left[(3-\mu^{2}) \frac{v}{cr} 
 + \frac{1+\mu^{2}}{c}\frac{\pa v}{\pa r} \right] I_{\nu}          \nonumber\\ 
&=& \kappa_{\nu}B _{\nu} - \kappa_{\nu} I_{\nu} -\sigma_{\nu} I_{\nu} 
 + \frac{1}{4\pi} \int \sigma_{\nu} I_{\nu} d \Omega,
\end{eqnarray}

\noindent
where \(\kappa_{\nu}\) and \(\sigma_{\nu}\) are the absorptive and
scattering opacities, respectively, \(B_{\nu}\) is the Planck function
and \(\mu\) is the cosine of the angle made by the radial direction
and the direction of the ray. This equation is solved numerically
using the Feautrier method with an approximate Lambda operator similar
to the one described by Hauschildt (1992).

To determine the gas temperatures, Equation (1) is solved
simultaneously with the energy equation and the first two moment
equations of Equation (1). The energy equation of the radiation plus
gas is written as

\begin{equation}
\frac{\pa}{\pa t}\left(e + \frac{E}{\rho}\right) = \epsilon -(P + f E)
\frac{\pa}{\pa t} \left( \frac{1}{\rho}\right) -4\pi \frac{\pa}{\pa
M_{r}} (r^{2} F) + (3f-1)\frac{v E}{\rho r},
\end{equation}
while the radiation energy and momentum equations are
\begin{equation}
\frac{\pa}{\pa t}\left(\frac{E}{\rho}\right) = 
\frac{c}{\rho}(\kappa_{P} a T^{4}-\kappa_{E} E) 
-f E \frac{\pa}{\pa t} \left( \frac{1}{\rho}\right)
-4\pi \frac{\pa}{\pa M_{r}} (r^{2} F) + (3f-1)\frac{v E}{\rho r}, 
\end{equation}
and
\begin{equation}
\frac{\pa F}{\pa t} =
-\left(c \chi_{F}+ \frac{2 v}{r}\right) F
-4\pi r^{2} \rho \left( c^{2} \frac{\pa (fE)}{\pa M_{r}} 
+ 2F \frac{\pa v}{\pa M_{r}}\right)-(3f-1)\frac{c^{2}E}{r}, 
\end{equation}
respectively, where $e$ is the thermal energy of ions and electrons
per unit mass, and $E$, $F$ and $f$ are the radiation energy density,
flux, and the Eddington factor defined as follows.

\begin{equation}
E = \frac{2 \pi}{c} \int_{0}^{\infty} d \nu \int_{-1}^{1} I_{\nu} d
\mu,
\end{equation}
\begin{equation}
F = 2 \pi \int_{0}^{\infty} d \nu \int_{-1}^{1} I_{\nu} \mu d \mu,
\end{equation}
\begin{equation}
f = \frac{\int_{0}^{\infty} d \nu \int_{-1}^{1} I_{\nu} \mu^{2} d \mu}
{\int_{0}^{\infty} d \nu \int_{-1}^{1} I_{\nu} d \mu}.
\end{equation}

Partial derivatives with respect to \(t\) in Equations (1)-(4) are all
Lagrangian time derivatives.  The absorptive and scattering parts of
the opacity are given as
\begin{equation}
\kappa_{\nu} = \epsilon (\kappa_{\rm b-b} + \kappa_{\rm b-f}) + 
\kappa_{\rm f-f},
\end{equation}
and 
\begin{equation}
\sigma_{\nu} = (1-\epsilon) (\kappa_{\rm b-b} + \kappa_{\rm b-f}) +
 n_{\rm e} \sigma_{\rm T},
\end{equation}

where $\kappa_{\rm b-b}, \kappa_{\rm b-f}$ and $\kappa_{\rm f-f}$ are
the bound-bound, bound-free, and free-free opacities, respectively;
$n_{\rm e}$ is the number density of free electrons, and $\sigma_{\rm
T}$ is the Thomson scattering cross section.  In the moment equations,
the energy mean $(\kappa_{E})$ and the Planck mean opacities
$(\kappa_{P})$ include only the absorptive part, while the flux mean
opacity $(\chi_{F})$ is the total opacity.

For bound-bound transitions, energy levels and transition
probabilities are taken from the compilation by Kurucz (1991).  For
bound-free data, we use the analytic fitting formula to the
photoionization cross sections given by Verner \& Yakovlev (1995).
Local Thermodynamic Equilibrium (LTE) is assumed to determine the
ionization balance and the level populations of each ion. However, the
non-LTE effect is approximately taken into account by assuming that
the value of the absorptive fraction $\epsilon$ is a constant less
than unity in Equations (8) and (9). Experiments of spectral syntheses
have shown that a value $\epsilon = 0.1$ is a reasonable choice for
this fraction (Baron et al. 1996).

It has been argued by several authors that the expansion would
increase the chance of interactions between radiation and matter
through line transitions, and thus the mean opacities become larger
than in static medium, especially at relatively early phases (Karp,
Lasher, \& Chan 1977; Eastman \& Pinto 1993; Blinnikov 1996).
However, we neglected this expansion effect in evaluating the mean
opacities, although this is one of the currently controversial issues.

The energy deposition due to the radioactive decays is calculated with
a one energy-group $\gamma$-ray-transfer code (Iwamoto 1997).  We
assume an absorptive opacity $\kappa_{\gamma} = 0.03$ and the complete
trapping of positrons.  The rest-frame flux is calculated from the
comoving-frame intensities using the following transformation:

\begin{equation}
F_{\nu, {\rm rest}} = 2 \pi \int_{-1}^{1}(\mu+\beta) I\left(\mu,
\frac{\nu}{1+\beta \mu}\right) d \mu,
\end{equation}
\begin{equation}
F_{\lambda, {\rm rest}} = \frac{\nu^{2}}{c} F_{\nu, {\rm rest}}.
\end{equation}

For the calculation of the light curves of CO60 and CO100 discussed in
the next subsection, we use about 200 radial mesh points to solve the
moment Equations (2)-(4), while 800 frequency and 50 radial mesh
points were used for the multi-frequency radiative transfer Equation
(1).

\subsection{Light Curve Models}

\begin{figure}
\centerline{\psfig{figure=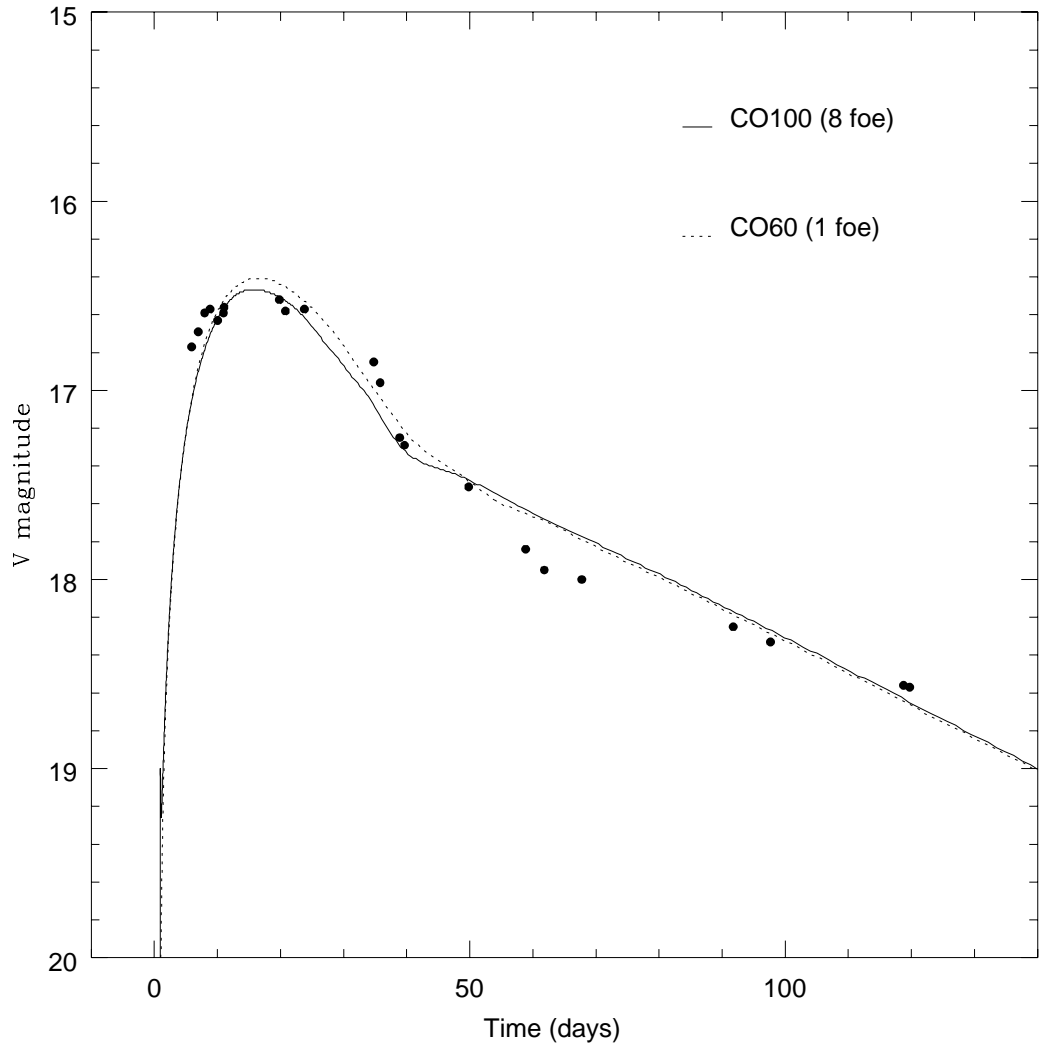,width=15cm}}
\caption{Calculated Visual light curves of CO60 and CO100 compared
with that of SN 1997ef.
\label{lc97ef}}
\end{figure}

In Figure \ref{lc97ef} we compare the calculated V light curves for
models CO60 and CO100 with the observed V light curve of SN1997ef.  We
adopt a distance of 52.3 Mpc (a distance modulus of $\mu=33.6$ mag) as
estimated from the recession velocity, 3,400 km s$^{-1}$ (Garnavich
\etal 1997a) and a Hubble constant $H_0=65$ km s$^{-1}$ Mpc$^{-1}$.
We assume no color excess $(E(B-V)=0.00)$; this is justified by the
fact that no signature of a narrow Na I D interstellar absorption line
is visible in the spectra of SN~1997ef at any epochs (Garnavich
1997a).  The light curve of SN 1997ef has a very broad maximum, which
lasts for $\sim$ 25 days.  This is much broader than in both the
ordinary SN Ic 1994I and the hypernova SN 1998bw.  The light curve
tail of SN~1997ef starts only $\sim 40$ days after maximum, much later
than in other SNe Ic.

The light curve of SN 1997ef can be reproduced basically with various
explosion models with different energies and masses.  In general, the
properties of the light curve are characterized by the decline rate in
the tail and the peak width, $\tau_{\rm peak}$.  The peak width scales
approximately as
\begin{equation}
\tau_{\rm peak} \propto \kappa^{1/2} M_{\rm ej}^{3/4} E_{\rm
K}^{-1/4},
\end{equation}
where $\kappa$ denotes the optical opacity (Arnett 1996).  This is the
time-scale on which photon diffusion and hydrodynamical expansion
become comparable.  Since the model parameters of CO100 and CO60 give
similar $\tau_{\rm peak}$, the light curves of the two models look
similar: both have quite a broad peak and reproduce the light curve of
SN1997ef reasonably well (Fig. \ref{lc97ef}).

The light curve shape depends also on the distribution of $^{56}$Ni,
which is produced in the deepest layers of the ejecta. More extensive
mixing of \ni\ leads to an earlier rise of the light curve.  For SN
1997ef, the best fit is obtained when the $^{56}$Ni is mixed almost
uniformly to the surface for both models.  Without such extensive
mixing, the rise time to V $=$ 16.5 mag would be $\sim$ 30 d for
CO100, which is clearly too long to be compatible with the
spectroscopic dating (see \S4).

Model CO60 has the same kinetic energy ($E_{\rm K} = 1$ \e{51} erg) as
model CO21, which was used for SN Ic 1994I (see Table 1 for the model
parameters). Since the light curve of SN 1997ef is much slower than
that of SN 1994I, the ejecta mass of CO60 is $\sim$ 5 times larger
than that of CO21.

The ejecta mass of CO100 is a factor of $\sim 2$ larger than that of
CO60, and it is only $\sim 20$\% smaller than that of model CO138,
which was used for SN 1998bw (Table 1).  Thus the explosion energy of
CO100 should be $\sim 8$ times larger than that of CO60 to reproduce
the light curve of SN 1997ef. This explosion is very energetic, but
still much weaker than the one in CO138.  The smaller $E_{\rm K}$ for
a comparable mass allows CO100 to reproduce the light curve of SN
1997ef, which has a much broader peak than that of SN 1998bw.

The light curve of SN 1997ef enters the tail around day 40.  Since
then, the observed V magnitude declines linearly with time at a rate
of $\sim 1.1 \times 10^{-2}$ mag day$^{-1}$, which is slower than in
other SNe Ic and is even close to the $^{56}$Co decay rate $9.6 \times
10^{-3}$ mag day$^{-1}$.  Such a slow decline implies much more
efficient $\gamma$-ray trapping in the ejecta of SN 1997ef than in SN
1994I.  The ejecta of both CO100 and CO60 are fairly massive and are
able to trap a large fraction of the $\gamma$-rays, so that the
calculated light curves have slower tails compared with CO21.

However, the light curves for both models decline somewhat faster in
the tail than the observations.  A similar discrepancy has been noted
for the Type Ib supernovae (SNe Ib) 1984L and 1985F (Swartz \& Wheeler
1991; Baron, Young, \& Branch 1993).  The late time light curve
decline of these SNe Ib is as slow as the \co\ decay rate, so that the
inferred value of $M$ is significantly larger (and/or $E_{\rm K}$ is
smaller) than those obtained by fitting the early light curve shape.
Baron et al. (1993) suggested that the ejecta of these SNe Ib must be
highly energetic and as massive as $\sim$ 50 \ms.  In \S5.1, we will
suggest that such a discrepancy between the early- and late-time light
curves might be an indication of asphericity in the ejecta of SN
1997ef and that it might be the case in those SNe Ib as well.

\subsection{Photospheric Velocities}

As we have shown, light curve modeling provides direct constraints on
$M_{\rm CO}$ and $E_{\rm K}$.  However, it is difficult to distinguish
between the ordinary SN Ic and the hypernova model from the light
curve shape alone, since models with different values of $M_{\rm ej}$
and $E_{\rm K}$ can reproduce similar light curves.  However, these
models are expected to show different evolutions of the photospheric
velocity and the spectrum as will be discussed in the following
sections.

\begin{figure}
\centerline{\psfig{figure=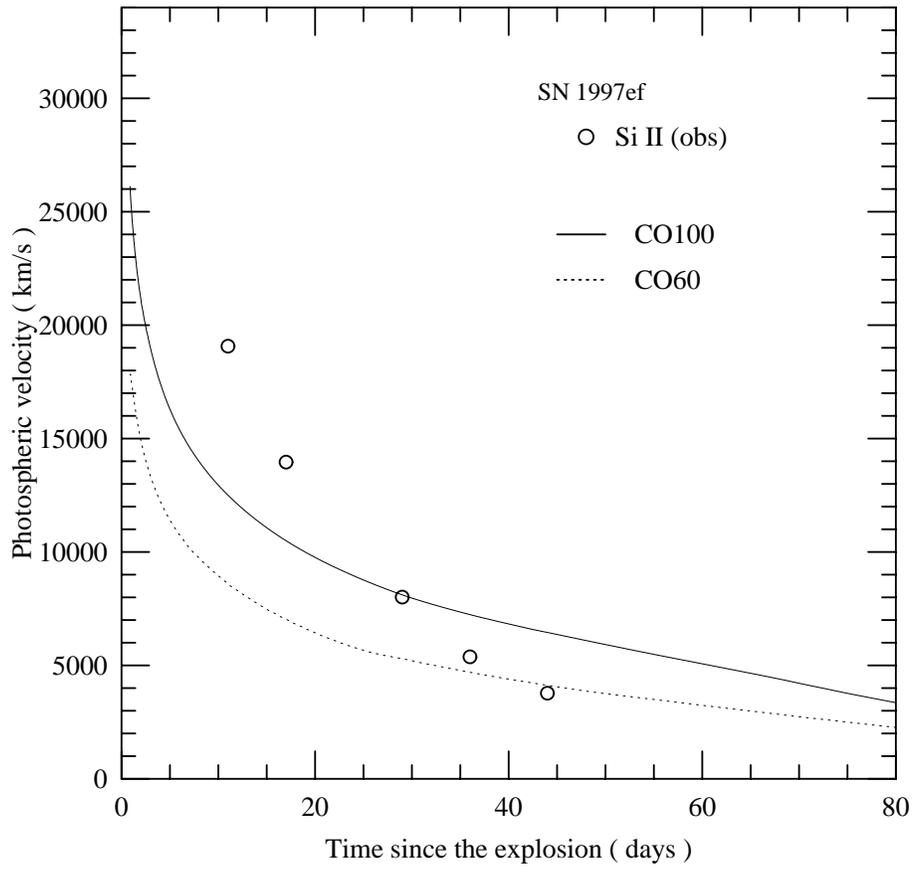,width=15cm}}
\caption{Evolution of the calculated photospheric velocities of
CO60 and CO100 (solid lines) compared with the observed velocities of the 
Si II 634.7, 637.1 nm line measured in the spectra at the absorption core.
\label{vph97ef}}
\end{figure}

The photospheric velocity scales roughly as $v_{\rm ph} \propto M_{\rm
ej}^{-1/2} E_{\rm K}^{1/2}$, so that $M_{\rm ej}$ and $E_{\rm K}$ can
be constrained by $v_{\rm ph}$ in a different way from by means of the
light curve width.  Figure \ref{vph97ef} shows the evolution of the
observed velocities of the Si~II line measured in the spectra at the
absorption core, and the velocities at the grey photosphere computed
by the light curve code for models CO60 and CO100.  The velocities of
the Si~II line are somewhat higher than that of the photosphere,
reaching $\sim$ 20,000 \kms at the earliest time.

In model CO60 the photosphere forms at velocities much smaller than
those of the observed lines, while CO100 gives photospheric velocities
as high as the observed ones.  It is clear, from this comparison, that
the hyper-energetic model CO100 is preferable to the ordinary model
CO60.  The apparent discrepancy that still exists between the CO100
and observations might be related to the morphology of the ejecta,
i.e., its deviation from spherical symmetry, as was also suggested in
the case of SN~1998bw (H\"oflich, Wheeler, \& Wang 1999; Iwamoto et al. 
1998). This issue will be discussed in \S5.1.

\section{SYNTHETIC SPECTRA} 

To strengthen the arguments in \S3.3, we compare the observed spectra
with theoretical model spectra computed using our explosion models
with a more sophisticated spectrum synthesis code(Mazzali \& Lucy
1993; Lucy 1999; Mazzali 1999).  With such a detailed spectrum
synthesis, we can distinguish between different models more clearly,
because the spectrum contains much more information than a single-band
light curve.

Around maximum light, the spectra of SN~1997ef show just a few very
broad features, and are quite different from those of ordinary SNe
Ib/c, but similar to SN~1998bw.  However, at later epochs the spectra
develop features that are easy to identify, such as the Ca~II IR
triplet at $\sim 8200$\AA, the O~I absorption at 7500 \AA, several
Fe~II features in the blue, and they look very similar to the spectrum
of the ordinary SN Ic 1994I.

We computed synthetic spectra with a Monte Carlo spectrum synthesis
code using the density structure and composition of the hydrodynamic
models CO60 and CO100.  The code is based on the pure scattering code
described by Mazzali \& Lucy (1993), but has been improved to include
photon branching, so that the reprocessing of the radiation from the
blue to the red is followed more accurately and efficiently (Lucy
1999; Mazzali 1999).

We produced synthetic spectra for three epochs near maximum, of
SN~1997ef: Nov 29, Dec 5, and Dec 17. These are early enough that the
spectra are very sensitive to changes in the kinetic energy.  As in
the light curve comparison, we adopted a distance modulus of
$\mu=33.6$ mag, and $E(B-V)=0.00$. The model parameters, the computed
temperatures and the magnitudes of the synthetic spectra for CO100 are
listed in Table 3.

\begin{figure}
\centerline{\psfig{figure=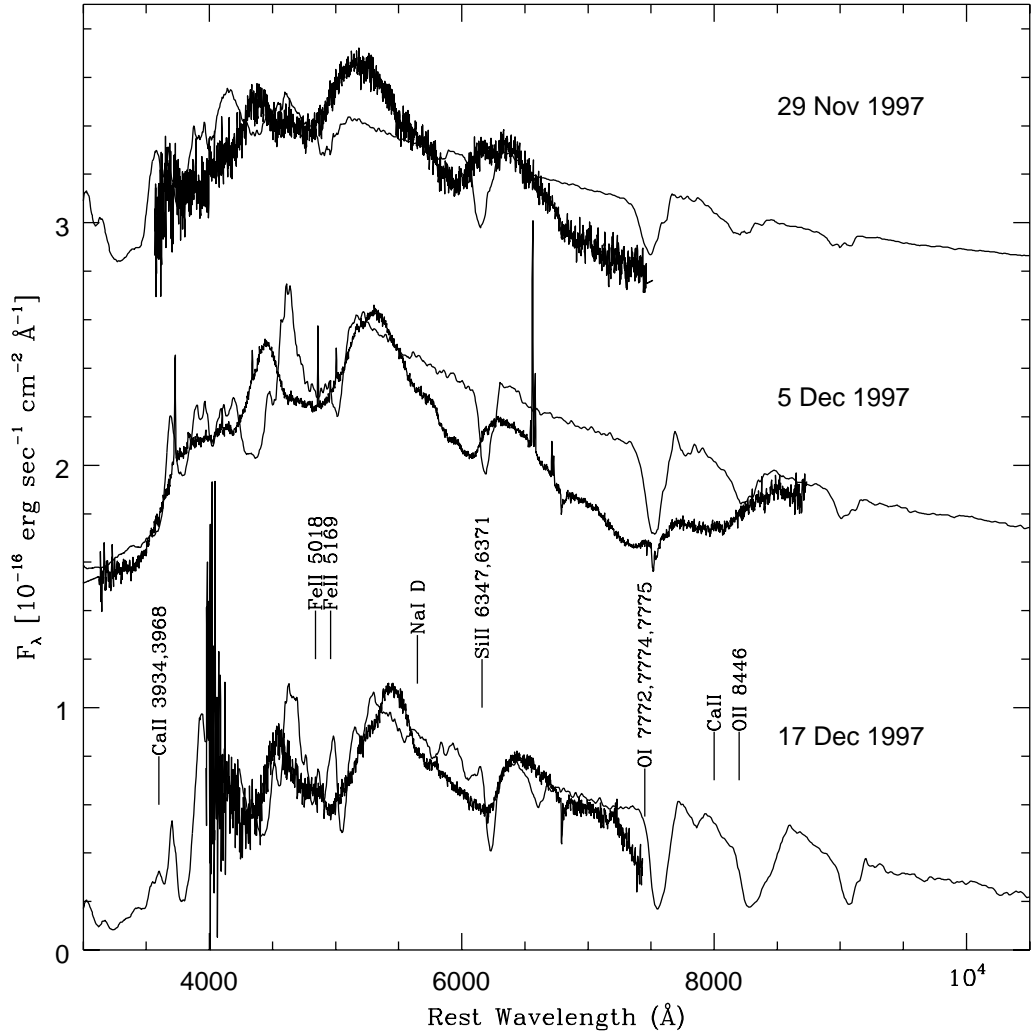,width=15cm}}
\caption{Observed spectra of SN~1997ef (bold lines) and synthetic
spectra computed using model CO60. The line features seen in the
synthetic spectra are much too narrow compared with observations.
\label{spco60}}
\end{figure}

In Figure \ref{spco60} we show the synthetic spectra computed with the
ordinary SN~Ic model CO60.  The lines in the spectra computed with
this model are always much narrower than the observations.  This
clearly indicates a lack of material at high velocity in model CO60,
and suggests that the kinetic energy of this model is much too small.

\begin{figure}
\centerline{\psfig{figure=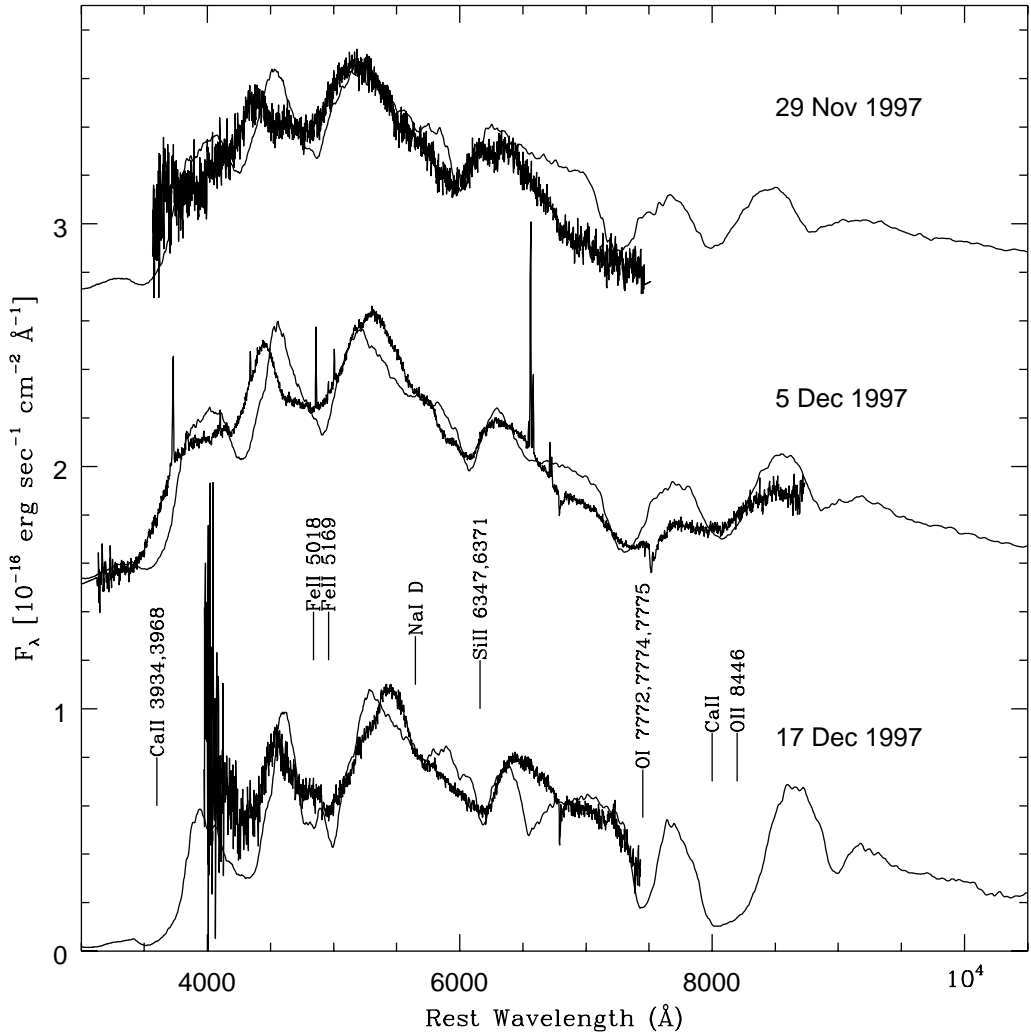,width=15cm}}
\caption{Comparison between the observed spectra of SN~1997ef (bold
lines) and synthetic spectra computed using model CO100 (fully drawn
lines). The fits are much improved with CO100 compared with the ones
with CO60.
\label{spco100new}}
\end{figure}

Synthetic spectra obtained with the hypernova model CO100 for the same
3 epochs are shown in Figure \ref{spco100new}.  The spectra show much
broader lines, and are in good agreement with the observations.  In
particular, the blending of the Fe lines in the blue, giving rise to
broad absorption troughs, is well reproduced, and so is the very broad
Ca-O feature in the red.  The two `emission peaks' observed at $\sim
4400$ and 5200\AA\ correspond to the only two regions in the blue that
are relatively line-free.  A similar situation is observed in
SN~1998bw (Iwamoto et al. 1998).

The spectra are characterized by a low temperature, even near maximum,
because the rapid expansion combined with the relatively low
luminosity (from the tail of the light curve we deduce that SN~1997ef
produced about $0.15 M_\odot$ of $^{56}$Ni, compared to about $0.6
M_\odot$ in a typical SN~Ia and $0.7 M_\odot$ in SN~1998bw) leads to
rapid cooling.  Thus the SiII 6355\AA\ line is not very strong.

Although model CO100 yields rather good synthetic spectra, it still
fails to reproduce the observed large width of the O~I - Ca~II feature
in the only near-maximum spectrum that extends sufficiently far to the
red (5 Dec 1997). An improvement can be obtained by introducing an
arbitrary flattening of the density profile at the highest velocities.

Full details of the spectrum synthesis calculations, including
insights on the density structure and the abundance in the ejecta will
be given in a separate paper (Mazzali et al., 1999, in preparation).

\section{DISCUSSION}

\subsection{Possible Aspherical Effects}

We have shown that the light curve, the photospheric velocities, and
the spectra of SN~1997ef are better reproduced with the
hyper-energetic model CO100 than with the ordinary SN Ic model
CO60. However, there remain several features that are still difficult
to explain with model CO100.

(1) The observed velocity of Si II decreases much more rapidly than
models predict. It is as high as $\sim$ 30,000 \kms\ at the earliest
phase, but it gets as low as $\sim$ 3,000 \kms\ around day 50
(Fig. \ref{vph97ef}).  We find that it is difficult to get such a
rapid drop of the photospheric velocity not only in models CO100 and
CO60, but also in other models that can reproduce the light-curve
shape reasonably well.  Models with higher energies and/or smaller
masses would be able to reproduce the fast evolution of the
photospheric velocity, but such models would inevitably produce light
curves with a narrower peak and a faster tail.

(2) Obviously, the observed light curve declines slower than model
CO100 in the tail part, and it is also a bit flatter than the model
near the maximum part(Fig. \ref{lc97ef}).  Models with lower energies
and/or larger masses are able to give improved fits to both the peak
and the tail of the light curve.  But, then it gets very difficult to
reproduce the large photospheric velocities observed at early times in
SN~1997ef.

This dilemma might be overcome if we introduce multiple components of
the light curve from different parts of ejecta moving at different
velocities.  In fact, the discrepancies may be interpreted as a
possible sign of asphericity in the ejecta: A part of ejecta moves
faster than average to form the lines at such high velocities at early
phases, while the other part of ejecta expands with a lower velocity
so that the low-velocity Si II line comes up at later epochs. Having a
low-velocity component would also make it easier to reproduce the slow
tail.

(3) Extensive mixing of $^{56}$Ni is required to reproduce the short
rise time of the light curve.  According to hydrodynamical simulations
of the Rayleigh-Taylor instability in the ejecta of envelope-stripped
supernovae (Hachisu \etal 1991; Iwamoto \etal 1996), large scale
mixing can not be expected to occur in massive progenitors, because in
the core of such massive stars the density gradient is not steep
enough around the composition interfaces.  One possibility to induce
such mixing in the velocity space is an asymmetric explosion(e.g.,
Nagataki, Shimizu, \& Sato 1998).  Higher velocity \ni\ could reach
the ejecta surface so that the effect of radioactive heating comes up
as early as is required from light curve modeling.

In order to realize higher densities at low velocity regions without
increasing the mass of ejecta significantly, it may be necessary that
the explosion is somewhat aspherical.  If the explosion is aspherical,
the shock would be stronger and the material would expand at a larger
velocity in a certain direction, while in its perpendicular direction,
on the other hand, the shock would be weaker, ejecting lower-velocity
material (e.g., H\"oflich et al. 1999).  The density of the central
region could be high enough for $\gamma$-rays to be trapped even at
advanced phases, thus giving rise to a slowly declining tail (see
Nakamura et al. 1999a for a discussion of SN~1998bw).  In the
extremely asymmetric cases, the material ejection may happen in a
jet-like form.  A jet could easily bring some \ni\ from the deepest
layer to the surface of high velocity.  Detailed spectral analysis of
observed spectra for different epochs are necessary to investigate
this issue further.

\subsection{Gamma-Ray Bursts/Supernovae Connection and SN 1997ef}

There have been an increasing number of candidates for the gamma-ray
burst(GRB)/supernova connection, including GRB980425/SN1998bw (Galama
et al.  1998; Iwamoto et al. 1998; Iwamoto 1999), GRB970514/SN1997cy
(Germany et al. 1999; Turatto et al. 1999), GRB980910/SN1999E
(Thorsett \& Hogg 1999).  Two other high-z GRBs may also be associated
with supernovae: GRB980326 (Bloom et al. 1999) and GRB970228 (Reichart
1999; Galama et al. 1999).  The optical transients of these GRBs
showed significant reddening and temporal slow down (even with a
second maximum) in their late light curves, which can be fitted by the
early power-law decay plus the red-shifted light curve of SN 1998bw.

As noted in \S1, a possible connection between SN 1997ef and GRB971115
has been suggested (Wang \& Wheeler 1998).  Recently another SN Ic,
1998ey, showed a spectrum with very broad features, very similar
spectra to that of SN1997ef on Dec 17 (Garnavich et al. 1998); but no
GRB counterpart has been proposed for SN~1998ey. Although this may
cast some doubt on the general association between hypernovae and
GRBs, it must be noted that both SNe 1997ef and 1998ey were less
energetic events than SN~1998bw. It is possible that a weaker
explosion is less efficient in collimating the $\gamma$-rays to give
rise to a detectable GRB (GRB980425 was already quite weak in
gamma-rays compared to the average GRBs), or that some degree of
inclination of the beam axis to the line of sight results in a
seemingly weaker supernova and in the non-detection of a GRB. Only the
accumulation of more data will allow us to address these questions.

\subsection{The Mass of Ejected \(^{56}\)Ni }

For the study of the chemical evolution of galaxies, it is important
to know the mass of $^{56}$Ni, $M(^{56}$Ni), synthesized in
core-collapse supernovae as a function of the main-sequence mass
$M_{\rm ms}$ of the progenitor star (e.g., Nakamura et al. 1999b).
From our analysis of SN 1997ef, we can add a new point on this
diagram.

We evaluate the uncertainty in our estimates of $M(^{56}$Ni) and
$M_{\rm ms}$. We need 0.15 \ms\ of $^{56}$Ni to get a reasonable fit
to the light curve of SN 1997ef at a distance $D = 52.3$ Mpc.  The
expected 10\% uncertainty in the distance leads to a 20\% uncertainty
in the $^{56}$Ni mass, i.e., $M(^{56}$Ni) $= 0.15 \pm$ 0.03 \ms.  The
distribution of $^{56}$Ni affects the peak luminosity somewhat, but
the effect is found to be much smaller than that of the uncertainty in
the distance.  A 10 \ms\ C+O star corresponds to a $M_{\rm ms} =$ 30 -
35 \ms, but the uncertainty involved in the conversion of the core
mass to $M_{\rm ms}$ may involve a larger uncertainty if the
progenitor undergoes a close binary evolution.

\begin{figure}
\centerline{\psfig{figure=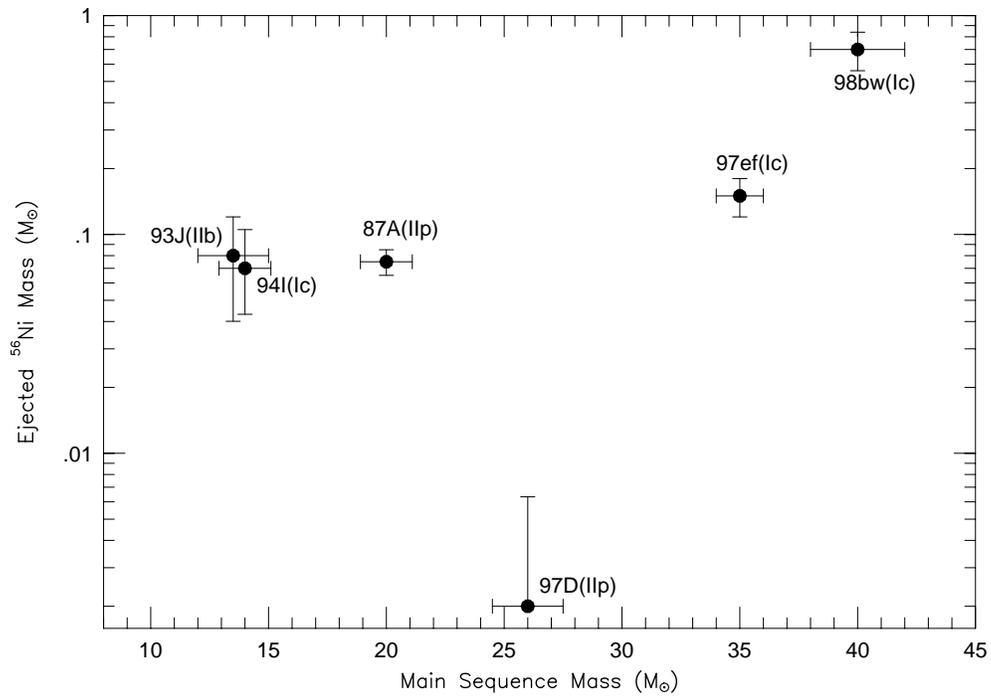,width=15cm}}
\caption{Ejected $^{56}$Ni mass versus the main sequence mass of the 
progenitors of several bright supernovae obtained from light curve models.
\label{nimass}}
\end{figure}

Figure \ref{nimass} shows $M(^{56}$Ni) against $M_{\rm ms}$ obtained
from fitting the optical light curves of SNe 1987A, 1993J, and 1994I
(e.g., Shigeyama \& Nomoto 1990; Nomoto et al. 1993, 1994; Shigeyama
\etal 1994; Iwamoto \etal 1994; Woosley et al. 1994; Young, Baron, \&
Branch 1995).  The amount of $^{56}$Ni appears to increase with
increasing $M_{\rm ms}$ of the progenitor, except for SN II 1997D
(Turatto et al. 1998).

This trend might be explained as follows.  Stars with $M_{\rm ms}
\lsim$ 25 \ms\ form a neutron star, producing $\sim$ 0.08 $\pm$ 0.03
\ms\ \ni\ as in SN IIb 1993J, SN Ic 1994I, and SN 1987A (although SN
1987A may be a borderline case between neutron star and black hole
formation). Stars with $M_{\rm ms} \gsim$ 25 \ms\ form a black hole
(e.g., Ergma \& van den Heuvel 1998); whether they become hypernovae
or ordinary SNe II may depend on the angular momentum in the
collapsing core.  For SN 1997D, because of the large gravitational
potential, the explosion energy is so small that most of \ni\ fell
back onto a compact star remnant; the fall-back might cause the
collapse of the neutron star into a black hole.  The core of SN II
1997D might not have a large angular momentum, because the progenitor
had a massive H-rich envelope so that the angular momentum of the core
might have been transported to the envelope possibly via a
magnetic-field effect.  Similarly, a negligible amount of ejection of
\ni\ in black hole formation has recently been suggested for X-ray
Nova Sco (GRO J1655-40), where the companion star of the black hole
seems to be enriched with S, Si, Mg, and O but not Fe (Israelian et
al. 1999).  Hypernovae such as SNe 1998bw, 1997ef, and 1997cy might
have rapidly rotating cores owing possibly to the spiraling-in of a
companion star in a binary system.  The outcome certainly depends also
on mass-loss rate and binarity.

As noted in \S5.2, it has been claimed that the optical afterglows of
GRB's 980326 and 970228 are better reproduced if a red-shifted light
curve of SN 1998bw is superposed on a power-law light component (Bloom
et al. 1999; Reichart 1999; Galama et al. 1999).  A question arising
from these two examples is whether the supernovae associated with GRBs
have a uniform maximum luminosity, i.e., whether 0.7 \ms\ \ni\
production as in SN 1998bw is rather common or not.  However, the
present study of SN 1997ef shows that the \ni\ mass and thus intrinsic
maximum brightness of SN 1997ef is smaller than in SN 1998bw by a
factor of 4 - 5 (see the next subsection).  We certainly need more
examples for defining the luminosity function and the actual
distribution of masses of \(^{56}\)Ni produced in
supernovae/hypernovae.

\subsection{Possible Evolutionary Scenarios}

Here we classify possible evolutionary paths leading to C+O star
progenitors.  In particular, we explore the paths to the progenitors
that have rapidly rotating cores with a special emphasis, because the
explosion energy of hypernovae may be extracted from rapidly rotating
black holes(Blandford \& Znajek 1977).

(1) Case of a single star: If the star is as massive as $M_{\rm ms}
\gsim$ 40 \ms, it could lose H and He envelopes in a strong stellar
wind (e.g., Schaller \etal 1992).  This would be a Wolf-Rayet star.

(2) Case of a close binary system: Suppose we have a close binary
system with a large mass ratio. In this case, the mass transfer from
star 1 to star 2 inevitably takes place in a non-conservative way, and
the system experiences a common envelope phase where star 2 is
spiraling into the envelope of star 1.  If the spiral-in releases
enough energy to remove the common envelope, we are left with a bare
He star (star 1) and a main-sequence star (star 2), with a reduced
separation.  If the orbital energy is too small to eject the common
envelope, the two stars merge to form a single star (e.g., van den
Heuvel 1994).

(2-1) For the non-merging case, possible channels from the He stars to
the C+O stars are as follows (Nomoto, Iwamoto, \& Suzuki 1995).

(a) Small-mass He stars tend to have large radii, so that they can
fill their Roche lobes more easily and lose most of their He envelope
via Roche lobe overflow.

(b) On the other hand, larger-mass He stars have too small radii to
fill their Roche lobes.  However, such stars have large enough
luminosities to drive strong winds to remove most of the He layer
(e.g., Woosley, Langer, \& Weaver 1995).  Such a mass-losing He star
would corresponds to a Wolf-Rayet star.

Thus, from the non-merging scenario, we expect two different kinds of
SNe Ic, fast and slow, depending on the mass of the progenitor.  SNe
Ic from smaller mass progenitors (channel a) show faster light-curve
and spectral evolutions, because the ejecta become more quickly
transparent to both gamma-ray and optical photons. The slow SNe Ic
originate from the Wolf-Rayet progenitors (channels b and 1).  The
presence of both slow and fast SNe Ib/Ic has been noted by Clocchiatti
\& Wheeler (1997).

(2-2) For the merging case, the merged star has a large angular
momentum, so that its collapsing core must be rotating rapidly.  It
would lead to the formation of a rapidly rotating black hole from
which possibly a hyper-energetic jet could emerge.  If the merging
process is slow enough to eject H and He envelopes, the star would
become a rapidly rotating C+O star.  Such C+O stars are the candidates
for the progenitors of Type Ic hypernovae like SNe 1997ef and 1998bw.
If a significant amount of H-rich (or He) envelope remains after
merging, the rapidly rotating core would lead to a hypernova of Type
IIn possibly like SN 1997cy (or Type Ib).

\section{CONCLUSIONS}

We have shown that the photospheric velocities and the spectra of
SN~1997ef are much better reproduced by the hyper-energetic model
CO100 than by the ordinary SN Ic model CO60. The model parameters
determined for CO100 are $E_{\rm K} = 8$ \e{51} erg, $M_{\rm CO} = 10
M_\odot$ (which corresponds to the C+O core of a 30--35 $M_\odot$
star), and $M$(\ni)$ = 0.15$ \ms.  The compact star remnant of CO100
is as massive as $M_{\rm rem} \sim$ 2.4 \ms, thus possibly being a
black hole.  This high explosion energy would be extracted from the
rapidly rotating black hole.

For SN~1997ef, $M_{\rm CO}$, $M$(\ni) and $E_{\rm K}$ are all slightly
smaller than for SN~1998bw, but SN~1997ef can certainly be regarded as
a {\sl hypernova} in terms of the kinetic energy of explosion.
Therefore, we suggest that SNe 1997ef, 1998ey, and 1998bw form a new
class of hyper-energetic Type Ic supernovae, which we call hypernovae.
They are distinguished by their large kinetic energies, 8 - 60 times
larger than in ordinary supernovae.

The smaller line velocities at advanced phases and the flatter light
curve tail of SN~1997ef than the models predict may suggest the
presence of a low-velocity, relatively dense core, while its higher
line velocities at early phases imply the presence of a
yet-higher-velocity component of ejecta.  These are very difficult to
be reconciled with any spherically symmetric models, even with the
high-energy spherical model CO100.  This discrepancy between models
and the observations, as well as the extensive mixing of \(^{56}\)Ni
required to explain the early rise of the light curve, seems to
indicate that the explosion of SN 1997ef was at least somewhat
aspherical.

This work was started at the Institute for Theoretical Physics,
University of California, Santa Barbara, USA, supported under NSF
grant no. PHY74-07194.  We would like to thank Drs. Hideyuki Umeda,
David Branch, and Nobert Langer for informative and stimulating
discussion. We would also like to thank the anonymous referee for his
or her useful comments and suggestions, which helped us improve the
contents of the paper.  This work has been supported in part by the
grant-in-Aid for Scientific Research (05242102, 06233101) and COE
research (07CE2002) of the Ministry of Education, Science, and Culture
in Japan, and the fellowship of the Japan Society for the Promotion of
Science for Japanese Junior Scientists (6728).  Part of the
computation was carried out on Fujitsu VPP-500 at the Institute of
Physical and Chemical Research (RIKEN) and at the Institute of Space
and Astronautical Science (ISAS).

\newpage

\newpage

\newpage

\begin{table}[t]
\tablenum{1}
\begin{center}
\centerline{Table 1.~Parameters of the CO star models}
\vspace*{4mm}
\begin{tabular}{ccccccc}
\hline model & C+O core mass~($M_\odot$) & ejecta mass~($M_\odot$) &
 \(^{56}\)Ni mass ($M_\odot$) & $E_{\rm K}$ (10$^{51}$ erg)  \\
\hline CO21  &  2.1 &  0.9 & 0.07 & 1  \\ 
\hline CO60  &  6.0 &  4.4 & 0.15 & 1  \\ 
\hline CO100 & 10.0 &  7.6 & 0.15 & 8  \\ 
\hline CO138 & 13.8 & 10.8 & 0.7  & \(\sim 30\) \\ 
\hline
\end{tabular}
\end{center}
\end{table}

\begin{table}[t]
\tablenum{2}
\begin{center}
\centerline{Table 2.~Predicted yields of SN1997ef ($M_\odot$)}
\vspace*{4mm}
\begin{tabular}{ccccccc} \hline
model& C & O & Si & S & Ca & Fe\\\
CO60& 5.2$\times 10^{-2}$ & 3.0 & 0.10 & 3.7$\times 10^{-2}$ &
5.7$\times 10^{-3}$ & 0.16 \\\
CO100& 0.58 & 5.6 & 0.42 & 0.19 & 2.5$\times 10^{-2}$ & 0.19 \\
\hline\hline
model& $^{44}$Ti & $^{56}$Ni & $^{57}$Ni \\
CO60 & 2.1$\times 10^{-4}$ & 0.15 & 5.7$\times 10^{-3}$ \\
CO100& 4.5$\times 10^{-5}$ & 0.15 & 5.7$\times 10^{-3}$ \\
\hline
\end{tabular}
\end{center}
\end{table}

\begin{deluxetable}{rcccccccccccccc}
\scriptsize
\tablenum{2}
\tablecaption{Parameters of the synthetic spectra}
\tablehead{\colhead{date} &
\colhead{epoch} &
\colhead{$L$} &
\colhead{$v_{ph}$} &
\colhead{$v_{SiII}$} & 
\colhead{$\log \rho_{ph}$} & 
\colhead{Mass} & 
\colhead{$T_{eff}$} &
\colhead{$T_{bb}$} &
\colhead{$B_{mod}$} &
\colhead{$V_{mod}$} &
\colhead{$V_{obs}$} &
\colhead{$BC$} &
\colhead{$M_{mod}$} \nl
\colhead{ } &
\colhead{days} & 
\colhead{erg s$^{-1}$} & 
\colhead{km s$^{-1}$} & 
\colhead{$R_{\odot}$} & 
\colhead{g cm$^{-3}$} &
\colhead{\ms} & 
\colhead{K} &
\colhead{K} & & & & &  \nl} 
\startdata
29 Nov &  9 & 42.17 & 15500 & 19072 & -12.65 & 0.71 & 6123 & 7666 & 
 17.45 & 16.75 & 16.7 & 0.28 & -16.700 \nl
 5 Dec & 15 & 42.19 &  9500 & 13962 & -12.30 & 3.02 & 6128 & 9407 & 
 17.35 & 16.63 & 16.5 & 0.35 & -16.750 \nl
17 Dec & 27 & 42.24 &  7500 &  8011 & -12.67 & 4.79 & 5291 & 6697 & 
 17.70 & 16.59 & 16.6 & 0.26 & -16.875 \nl
\enddata
\end{deluxetable}

\end{document}